\documentstyle[epsfig,11pt,amssymb,amsmath]{article}

\def\be{\begin{eqnarray}}
\def\ee{\end{eqnarray}}
\def\ba{\begin{array}}
\def\ea{\end{array}}

\def\ie{{\it i.e. }}
\def\cf{{\it cf }}
\def\vev{{\it vev }}
\def\tr{\mbox{tr }}
\def\Imm{\Im m \, }
\def\Ree{\Re e \, }
\def\tr{{\mbox{tr}\,}}
\def\dim{{\mbox{dim}\,}}
\def\rank{{\mbox{rank}\,}}
\def\diff#1#2{{\frac{\partial #1}{\partial #2 }}}
\def\l|{\left| \vphantom{\bigl(}\right. }
\def\r|{\left. \vphantom{\bigl(} \right|}
\def\//{/\hspace{-.1cm}/}

\begin{document}
\begin{titlepage}

\begin{flushright}
CERN-TH/97-171\\
Saclay T97/084 \\
hep-th/9707226 \\
\end{flushright}

\vskip.5cm
\begin{center}
{\huge{\bf Geometrical approach to duality in $N=1$ 
supersymmetric theories}}
\end{center}
\vskip1.5cm

\centerline{ Ph. {\sc Brax} $^{a}$, C. {\sc Grojean} $^{a,b}$
{\it and} 
C.A. {\sc Savoy} $^{a,b}$}
\vskip 15pt
\centerline{$^{a}$ CEA-SACLAY, Service de Physique Th\'eorique}
\centerline{F-91191 Gif-sur-Yvette Cedex, {\sc France}}
\vskip 3pt
\centerline{$^{b}$ CERN-TH}
\centerline{CH-1211 Geneva 23, {\sc Switzerland}}
\vglue .5truecm

\begin{abstract}
We investigate the geometry of the moduli spaces of dual electric and magnetic 
$N=1$ supersymmetric field theories. Using the $SU(N_c)$ gauge group as 
a guideline we show that the electric and magnetic moduli spaces 
coincide for a suitable choice of the K\"ahler potential of the magnetic theory.
We analyse the K\"ahler structure of the dual moduli spaces. 
\end{abstract}



\vfill
\begin{flushleft}
CERN-TH/97-171\\
July 1997\\
\end{flushleft}

\vfill
\footnoterule
\noindent
e-mail addresses:\\
brax, grojean, savoy@spht.saclay.cea.fr

\end{titlepage}


The long-distance behaviour of $N=1$ supersymmetric field theories is believed 
to unravel a new kind of duality \cite{duality}.  Two theories described in the
ultraviolet by  different gauge contents and gauge groups flow in the infrared
towards  the same  fixed point of the renormalization group.   They effectively
describe the same long-range physics.   The non-trivial feature of this duality
is that the gauge groups of  the dual pairs are different.   The gist of this
relationship is better understood in terms of  the dual pair of moduli
spaces.    In general the scalar potential of a supersymmetric gauge theory
comprises  two terms known as the $D$- and $F$- terms.   The possible
supersymmetric vacua are  labelled by the vanishing of both the $D$- and $F$-
terms.  The locus of  the scalar fields where the scalar potential vanishes is
the moduli space.  In general  the gauge group is  completely broken by these
vacua.   The moduli space is obtained when the spurious degrees of freedom
associated to the Higgs mechanism are eliminated.       The structure of the
classical moduli space is preserved to  all orders in perturbation by the
non-renormalization theorem.   In fact non-perturbative effects do not modify
the moduli space of dual pairs  in all the known examples. This implies that the
low-energy  supersymmetric field theory is a supersymmetric non-linear sigma
model  whose target space is the (classical) moduli space. 

When there is no superpotential, the moduli space reduces to the 
equations $D^A=0$, where $A$ labels the generators of the gauge group $G$. 
The moduli space is the quotient $(D^A=0)/G$ after going to 
the unitary gauge. 
It is a subset of the (flat) vector space $V$ spanned by the scalar fields 
of the ultraviolet theory.

The key idea for the construction of the dual magnetic theory stems from  a deep
result in geometric-invariant theory. The moduli space is  homeomorphic to
the space of closed orbits in $V$ under the action of the  complexified gauge
group $G^c$. This allows the identification of  the chiral ring of  
the electric theory
with the ring of  $G$-invariant polynomials \cite{duality}.  The dual magnetic
theory corresponds to the same superconformal field theory  and is characterized
by the same chiral ring as the electric theory.  
{}From a geometric point of
view, the equality between the chiral rings of  the electric and magnetic
theories implies that the electric and  magnetic moduli spaces are isomorphic as
algebraic varieties.  

In the present paper we shall study the isomorphism of the
electric and  magnetic moduli spaces from the point of view of complex
K\"ahlerian geometry.  In particular the K\"ahlerian equivalence between the 
electric and  magnetic moduli space entails that the corresponding low-energy 
non-linear sigma models are identical.

\vspace{.2cm}
In supersymmetric theories, holomorphy is probably one of the most important
principles \cite{Shifman}--\cite{Seiberg-Witten}.
Usually, the scalar potential is a sum of $F$- and $D$-terms: the $F$-terms are
the norm of the gradient of a holomorphic function, the
superpotential $W$. For the $D$-terms, the relation with holomorphy is a bit
more subtle and appears when we are interested in supersymmetric vacua
\cite{Buccella}. Consider for instance a compact Lie group $G$ with Hermitian
generators $T^A$ acting on a representation containing some scalar fields $z$;
then
\be
D^A \equiv \tr \left(
{\diff{{\cal K}}{z}}
T^A z
\right)
\ee
where  ${\diff{{ \cal K}}{z}}$ denotes the gradient of the K\"ahler potential
${\cal K}$ with respect to the scalar fields $z$.

A sufficient and necessary condition for the vanishing of the $D$-terms was
given in \cite{Buccella} in terms of holomorphic gauge invariants: for any
holomorphic gauge-invariant polynomial $I$ in the scalar fields, each \vev
$\xi$ verifying 
\be
\diff{I}{z^{a}} _{\l| z=\xi} = C \diff{{\cal K}}{z^a} _{\l| z=\xi}\ ,
\label{invariant} 
\ee
where $C$ is a complex  constant, is a solution of the set of  equations
$D^A=0$ (of course, the whole $G$-orbit associated to such a  \vev is also a
solution), and to any solution of $D^A=0$ one can associate a  holomorphic
gauge  invariant satisfying (\ref{invariant}). The proof of this result
\cite{Gatto-Procesi} was obtained  by studying the closed orbits of the 
complexified
gauge group $G^c$ and the ring of $G$-invariant polynomials.
This ring is finitely generated: one can find an
integrity basis \ie a set of $G$-invariant holomorphic polynomials  
$\left\{ I^a (z) \right\}_{a=1\cdots d}$  
such that every $G$-invariant polynomial in $z$ can
be written as a polynomial in the $I^a (z)$. However the elements of  an
integrity basis are not always   algebraically independent, \ie there exist 
algebraic relations (called {\it syzigies}) satisfied by the fundamental 
invariants\footnote{A trivial example is provided by the $SU(N_c)$ gauge theory
with $N_c$ fields in the fundamental and antifundamental representations; the
fundamental invariants are the mesons $M=z{\tilde z}$ and the baryons $B=\det z$
and  ${\tilde B}= \det {\tilde z}$; classically, they are constrained by 
$\det M- B {\tilde B}=0$.}. 
Now  every $G$-invariant holomorphic polynomial $I$ is
automatically $G^c$-invariant, so that to each $G^c$-orbit 
corresponds a vector in
${\Bbb{C}}^d$ made out  of the values of fundamental invariants 
along this orbit.
Conversely, it can be shown that to each vector in ${\Bbb{C}}^d$ satisfying  
the
syzigies is associated a unique closed $G^c$-orbit. In that sense the algebraic
subset of ${\Bbb{C}}^d$ defined by the syzigies is  
identified with the set of closed $G^c$-orbits.

Equations (\ref{invariant}) can be seen  as a condition for the points of a
closed $G^c$-orbit to extremize the K\"ahler potential, \ie the Euclidean length
in flat space, the constant $C^{-1}$ being a Lagrange multiplier. 
A result of
geometric invariant theory \cite{Kempf} states that the  points extremizing the 
K\"ahler potential on a $G^c$-orbit form a unique $G$-orbit. 
Identifying   the
points on a same $G$-orbit,  there is a one-to-one correspondence between the
solutions of $D^A=0$ and the closed $G^c$-orbits.

We can also interpret  (\ref{invariant}) in another way: keep 
${\cal K}$ constant and extremize an invariant $I$ (now $C$ itself 
appears as a Lagrange multiplier). 
Here we work on an iso-K\"ahler surface and we analyse how
the value of the invariant changes from one $G^c$-orbit to another.

Yet, the description of the moduli space in terms
of  polynomial invariants is not well adapted to the calculation of the 
K\"ahler potential induced from a flat K\"ahler potential on $V$.  
In some simple examples \cite{ADS}, \cite{Poppitz-Randall}, the moduli space has
been described in terms of polynomial invariants and the corresponding K\"ahler
potential induced from the flat K\"ahler
potential to the flat directions. However, in these few examples, 
there are no syzigies to constrain 
the integrity basis, which is then very simple.
The syzigies are an obvious obstacle to any attempt to derive
the low-energy effective theory in terms of the analytic invariants  $I^a$, as
they have to be solved to define a set of independent analytic coordinates.
This corroborates the fact that the 't~Hooft anomaly matching conditions cannot 
be verified by using the polynomial invariants.  For all these reasons we do not
approach the question in terms of these invariants.

To identify and to count the moduli fields among the scalars in a given theory
is quite easy. Correspondingly to the $D$- and $F$-terms in the potential one
obtains mass matrices $M_D^2$ and $M_F^2$ for the scalars. If the gauge
group $G$ is broken down to $H$, the scalars along the  $\dim (G/H)$ complex
directions  $\langle  D^A_i \rangle z^i$ complete the massive vector multiplets
corresponding to the broken generators, so that (for $CP$-conserving vacua)
$CP$-even scalar components of these supermultiplets are the only eigenstates of
$M_D^2$ with non-zero eigenvalues. Their $CP$-odd partners are unphysical degrees
of freedom associated to the Goldstone zero eigenstates of $M_D^2$.  This
eliminates $2 \, \dim (G/H)$ scalars that are certainly not moduli. 

The gauge invariance of the superpotential entails, at a supersymmetric vacuum,
$ F_{ij} D^A_{{\bar \imath}} =0$, which explicitly shows that the massive
eigenstates of $M_F^2$ are orthogonal to the scalars absorbed in the massive
vector  multiplets. Hence one has to eliminate $\rank (F_{ij})$ complex scalars,
leaving $\dim V -2\, \dim (G/H) - 2\, \rank (F_{ij})$ 
massless states. Then one has
to remove all massless states with gauge interactions (non-singlet under the
unbroken $H$ subgroup) and superpotential interactions, involving only the
massless states. 
Of course this identification gives no further
information on the geometry of the moduli manifold.

In order to identify this manifold, one has to determine the symmetries of the
flat potential condition $V=0$, namely the simultaneous zeros of the $D^A$'s 
and $F_i$'s :
\be
&D^A (\xi, \xi^*) \equiv \xi^\dagger T^A \xi = 0 \ ;&\nonumber\\
&F_i (\xi) = 0 \ .&
\ee
These symmetries include of course non-compact transformations. For instance,
under a dilation plus overall phase transformation, $\delta \xi = \epsilon \xi$,
the invariance of the moduli space requires $F_{i,\xi} \xi =0$. 
The homogeneity of the superpotential in the fields is a sufficient, not
necessary in general, condition for the dilation invariance of the moduli space.
The identification of the complete symmetries of the vacuum degeneracy defines
the geometry of the moduli space. In particular, one must find the same moduli
spaces for dual supersymmetric theories.

\vspace{.2cm}
We shall carry out our analysis in the explicit context of the $SU(N_c)$  gauge
theory with $N_f$ flavours of quarks and antiquarks.  The classical moduli space
is not modified non-perturbatively when $N_f>N_c$.  The construction of the
duality between the electric and the magnetic theories  has been obtained in the
conformal range ${\frac{3}{2}}N_c<N_f<3 N_c$ in which the electric and magnetic
theories are both asymptotically free.

Let us first provide a geometrical description of the so-called electric
theory \cite{duality}. This theory possesses gauge invariance $SU(N_c)$ with
$N_f$ flavours of quarks and antiquarks transforming as the fundamental and
antifundamental representations. We denote by $Q^i_\alpha$ and 
${\tilde Q}^{\tilde \imath}_{\tilde \alpha}$ the corresponding scalar fields.  
The K\"ahler potential is 
\be
{\cal K} (Q, Q^\dagger, {\tilde Q} , {\tilde Q}^\dagger )
=\tr ( Q^\dagger Q + {\tilde Q}^\dagger {\tilde Q}) \ .
\label{K_electric} 
\ee
The
anomaly-free global symmetries of the theory are
$ SU(N_f) \times SU(N_f) \times U(1)_B \times U(1)_R $, 
with $Q^i_\alpha$ transforming as $(N_f,1,1,(N_f-N_c)/N_f)$ and ${\tilde
Q}^{\tilde \imath}_{\tilde \alpha}$ as $(1,{\bar N_f},-1,(N_f-N_c)/N_f)$.

The moduli space parametrizes the variety of degenerate supersymmetric vacuum
states, \ie states with zero energy. Here these vacua are solutions of 
the $D$-flatness
equations: 
\be
D^A \equiv (Q^i_\alpha)^* T^A_{\alpha\beta} Q^i_\beta
- {\tilde Q}^{\tilde \imath}_{\tilde \alpha} T^A_{{\tilde \alpha}{\tilde \beta}}
({\tilde Q}^{\tilde \imath}_{\tilde \beta})^*
&=& 0 \ ,
\label{D=0} 
\ee
where $A$ runs over all the gauge indices. 
The previous equations are equivalent to the single relation: 
\be
(Q^i_\alpha)^* Q^i_\beta 
- {\tilde Q}^{\tilde \imath}_{\tilde \alpha}
({\tilde Q}^{\tilde \imath}_{\tilde \beta})^*
= \lambda \, \delta_{\alpha\beta} \ ,
\label{Q=1}
\ee
where $\lambda$ is  a real number.

Equation (\ref{Q=1})  explicitly exhibits an $SU(N_f,N_f)\times U(1)_B$ flavour
invariance \cite{Buccella}, acting on the $2N_f$ component vectors 
$(Q\, , \, {\tilde Q}^* )$,
including a Cartan generator corresponding to the $U(1)_R$. Moreover there is
an obvious invariance under dilation. Finally the symmetry of the moduli
space  is $G_e \times SU(N_c)_G$, where 
\be
G_e = U(N_f,N_f) \times D  \ .
\ee
Notice that holomorphy is not preserved by the action of $G_e$.

Due to the constraint (\ref{Q=1}), possibly with the interchange of the role of
$Q$ and ${\tilde Q}$, and by making an 
$SU(N_f)\times SU(N_f) \times U(1)_B \times
U(1)_R \times SU(N_c)$ transformation, any solution of (\ref{D=0}) can be put
in  the form
\be
(Q^i_\alpha \, , \, 
{\tilde Q}^{\tilde \imath *}_{\tilde \alpha} )
& = & 
\left\{ 
\begin{array}{cl}
\left(\sqrt{a_i^2+\lambda^2}\, \delta_{\alpha}^i
\, , \, a_i \,\delta_{\alpha}^i \right) 
& \mbox{for } \ i=1 \cdots N_C \ ; \\
(0,0) & \mbox{for } \  i=N_c+1 \cdots N_f\ .
\end{array}
\right.
\label{vac_gene}
\ee
with $a_i$ real and positive. Actually  these solutions belong to two kinds of
orbits of $U(N_f,N_f)\times SU(N_c) \times D$, corresponding to the two  cases
where $\lambda$ is zero and not.

Consider first the case $\lambda \not = 0$ \footnote{There are two  conjugated
baryonic branches corresponding to  $\lambda>0$ and $\lambda<0$.
Each of them forms a single orbit under  the symmetry group. In the following we
deal with the case $\lambda >0$.  The case $\lambda<0$ is obtained by exchanging
the roles of $Q$ and  ${\tilde Q}^*$. }; then using the appropriate element in
the subgroup $U(1,1)^{N_c}$ of $U(N_f,N_f)$, one gets
\be
Q^i_{\alpha} = 
\left\{ 
\begin{array}{cl}
\delta^i_\alpha & \mbox{for } \  i=1\cdots N_c\ ;\\
0 & \mbox{for } \  i=N_c+1 \cdots N_f\ ;
\end{array}
\right.
&&
{\tilde Q}^{\tilde \imath}_{\tilde \alpha} = 0
\ .
\label{vac_baryon}
\ee
We shall refer to the orbit associated with this point as the time-like orbit or
{\it baryonic orbit}. When $\lambda=0$, the situation is  different since one
cannot use non-compact symmetries to cancel some components, but only to fix all
the non-vanishing values as follows
\be
Q^i_{\alpha} = 
{\tilde Q}^{\tilde \imath}_{\tilde \alpha} =
\left\{ 
\begin{array}{cl}
\delta^i_\alpha & \mbox{for } \ \alpha=1\cdots r\leq N_c \ ;\\
0 & \mbox{elsewhere} \ .
\end{array}
\right. 
\label{vac_meson}
\ee 
These orbits will be called {\it mesonic}. Therefore, the vacua manifold has
been identified as a finite set of $SU(N_f,N_f)\times U(1)_B \times SU(N_c)
\times D$ orbits. Each of these can be represented by the quotient of the
symmetry group (acting transitively on the orbit) by the stabilizer (or little
group) of one
point. Thus we have to identify the stabilizer of each representative point
considered earlier to characterize each orbit.

For the baryonic orbit, the little group associated to the vacuum
(\ref{vac_baryon}) takes a simple structure of direct product
\be
H_e = SU(N_f-N_c,N_f)\times U(1) \times SU(N_c)_D \ ,
\ee
where $SU(N_c)_D$ is the diagonal  combination of the gauge $SU(N_c)_G$ and an
$SU(N_c)$ subgroup of $U(N_f,N_f)$. The $U(1)$ is a combination of the $U(1)_B$
and an element of the  Cartan subalgebra of $SU(N_f,N_f)$. 
Here the gauge group is
completely broken; then after eliminating  the spurious massless scalars
associated to the Higgs mechanism, the real dimension of the coset $G_e/H_e$ is
$4N_fN_c - 2 N_c^2 +2 $. 

In order to further characterize the geometry of the baryonic branch of moduli
space, we first extract a flat subspace associated  to the diagonal
$Gl(1,{\Bbb{C}})$ factor in the coset and then introduce as coordinates the
$N_C\times (2N_f-N_c)$ complex  matrix $z$ and define its transformation under
$U(N_f,N_f)$ as follows.  First parametrize the Lie algebra of $U(N_f,N_f)$ in
the representation of dimension $2N_f$ by  
\be
\left(
\begin{array}{cc}
a & m \\
-\eta m^\dagger & -d 
\end{array}
\right) \ ,
\label{U(N,N)}
\ee
where $\eta$ is the $(N_f-N_c,N_f)$ signature, 
$a$ is an $N_c\times N_c$ anti-Hermitian matrix, $d$ is a
$(2N_f-N_c)\times (2N_f-N_c)$ matrix verifying $d \eta = - \eta d^\dagger$ and
$m$ is an $N_c\times (2 N_f-N_c)$ matrix. The action on the matrix $z$ takes the
non-linear infinitesimal form 
\be
\delta z = m + a z + z d + z \eta m^\dagger z \ .
\label{infinitesimal_action}
\ee
A finite transformation takes the homographic form
\be
z \to (Az+B) (Cz+D)^{-1} \ .
\label{homographic}
\ee
The elements of the  coset are then parametrized by the exponentials 
$e^{t(z)}$, where
\be
t(z)=
\left(
\begin{array}{cc}
0 & z \\
-\eta z^\dagger & 0 
\end{array}
\right)
\ .
\ee  
The action (\ref{infinitesimal_action}) of $U(N_f,N_f)$ on $z$ yields a 
non-linear action on the coset 
\be
g \ :  \ \ z \, \rightarrow \, z_g \ \ ,\ \ e^{t(z)} \, \rightarrow \, 
e^{t(z_g)} \ .
\ee
This action is transitive. There is only one $U(N_f,N_f)$-invariant
K\"ahler potential  up to a  K\"ahler transformation,
namely: ${\cal K} = \tr \ln (1+z \eta z^\dagger) .$

Let us now check the correspondence between the massless scalar fields and the
moduli fields. As already discussed previously, since there is no
superpotential and the whole gauge group is Higgsed around the vacuum
(\ref{vac_baryon}), all the scalars are moduli with the exception of the
$N_c^2-1$ complex scalars given by $T^A_{i\beta}  Q^i_\beta $, associated to
the $SU(N_c)$ massive vector multiplets.  All the $4N_fN_c-2N_c^2+2$ remaining
scalar fields are massless, and their number coincides with the real dimension
of $G_e/H_e$.

For the {\it mesonic orbit}, the pattern is more complicated since the little
group associated to the vacuum (\ref{vac_meson}) now has a structure of
semi-direct product, $H_r\times SU(N_c-r)_G$, with:
\be
H_r =  SU(N_f-r,N_f-r)\times U(1)^2\times SU(r)_D\otimes {\tilde H} \ ,
\label{stab_elec_meson}
\ee
where $SU(r)_D$ is the diagonal  combination of the  gauge and flavour $SU(r)$
subgroups,  the two $U(1)$'s are combinations of the $U(1)_B$, an element of
the  Cartan subalgebra of $SU(N_f,N_f)$ and an element of the Cartan subalgebra
of $SU(N_c)_G$. The semi-direct factor ${\tilde H}$ is a flavour nilpotent
subgroup  with generators transforming as  $(2(N_f-r), \overline{r}) \oplus
(\overline{2(N_f-r)},r)$ under  $U(N_f-r,N_f-r)\times SU(r)_D$ and the Abelian
subalgebra defined by their commutators, which transform as 
$(1, Adj\oplus 1)$ .

Notice that in this case, at least if $r < N_c-1$, an $SU(N_c-r)$ gauge 
subgroup remains unbroken. After eliminating the degrees of freedom 
moving to the massive vector multiplet, the  real dimension  of the coset
$G_e/H_r$ is $4N_fr-2 r^2$, which is the dimension of the moduli space. On
the mesonic orbit, the action of $SU(N_c)$ can be complexified; nevertheless,
the action of $U(N_f,N_f) \times SU(N_c)_G$ is already transitive. Let us check
this dimension from the massless spectrum around the vacuum  (\ref{vac_meson}).
As discussed in the previous section, starting from $4N_fN_c$ scalars, one has
to exclude $2(2N_cr-r^2)$ scalars, corresponding to the Higgsed part of
$SU(N_c)/SU(N_c-r)$, and $4(N_f-r)(N_c-r)$ massless scalars (corresponding to
$N_f-r$ flavours  of quarks and antiquarks), which are charged under the unbroken
gauge group $SU(N_c-r)$. Hence the remaining $4N_f r- 2 r^2$, 
\ie $\dim (G_e/H_r)$, ones are gauge
singlets.

The mesonic orbits are stratified by the index $r$.  The stratum corresponding
to  $r=N_c$ is such that the gauge group is completely broken.  In fact the
mesonic orbits correspond to the ``infinitely boosted'' baryonic orbit. Indeed
let us choose a point (\ref{vac_gene}) on the baryonic orbit,  \ie with
$\lambda\ne 0$,  and apply $r$  transformations in $U(1,1)$ of rapidities 
$\theta_i$ such that
\be
e^{\theta_i}\, 
\left( a_i^2+\sqrt {{\tilde a}_i^2+\lambda^2} \right)=e^{\theta} \ ,
\ee
where $ \theta$ goes to infinity. Then using the dilation invariance of  the
baryonic orbit, one can rescale the vacuum solution  by a common factor
$e^{-\theta}$.  This entails that the $r$ chosen directions are taken to unity
and  the remaining ones to zero.  The vacuum solution eventually tends to the
mesonic solution  (\ref{vac_meson}).  Hence the moduli space is the closure of
the baryonic orbit,  \ie by applying appropriate boosts and a global dilation
to  the baryonic orbit one can converge to  all the strata of the mesonic
orbits. 
In the same way, by applying infinite boosts, we can go from a mesonic
orbit with a stratification index $r$ to a more singular one with lower index.
Geometrically,
these boosts correspond to the shrinking of some circles in the moduli space. 
{}From a physical
point of view, the stratification index is  related to the number of massless
singlets of the theory.  As the stratification index goes  from $r$ to $r-1$,
the corresponding orbits differ by a dimension  $4(r-N_f)-2$.

\vspace{.2cm}
We now turn to the analysis of 
the dual theory, or magnetic theory. 
It possesses a gauge
group $SU(N_f-N_c)$ and contains $N_f$ chiral superfields, 
denoted $q_{\alpha}^i$ and 
$\tilde q_{\tilde \alpha}^{\tilde \imath}$, in the fundamental and 
antifundamental 
representations, together with a gauge-singlet matrix field 
$M^{i {\tilde \jmath}}$. The superpotential of the magnetic theory is
\cite{duality}
\be
W=\tr (q^iM^{i {\tilde \jmath}}\tilde q^{\tilde \jmath}) \ ,
\ee
and the K\"ahler potential is 
\be
{\cal K}= \tr (q^{\dagger} q) + 
\tr ( {\tilde q}^{\dagger} {\tilde q} ) + K (M^{\dagger}M) \ ,
\label{K_magnetic}
\ee
corresponding to the assumption of a flat metric for the quarks.
The K\"ahler manifold for the meson fields $M$ 
should be deduced from
the duality with the electric theory, as discussed later. 

The moduli space of the magnetic theory is defined as the set of 
possible vacua; 
as such, it is identified with the solutions of the equations:
\be
D^A & \equiv &
(q^i_{\alpha})^* T^A_{\alpha\beta} q_{\beta}^i 
- \tilde q^{\tilde \imath}_{\tilde\alpha} T^A_{\alpha\beta}
(\tilde q_{\tilde\beta}^{\tilde \imath})^*
=0 \nonumber \ ;\\
F_{ q^i_\alpha} & \equiv & 
M^{i {\tilde \jmath}} \tilde q_{\tilde\alpha}^{\tilde \jmath}
=0 \nonumber \ ;\\
F_{{\tilde q}^{\tilde \jmath}_\alpha} & \equiv &
q^{i}_{\alpha}M^{i {\tilde \jmath}}
=0\nonumber \ ; \\
F_{M^{i {\tilde \jmath}}} & \equiv & 
q^i_{\alpha}\tilde q_{\tilde \alpha}^{\tilde \jmath} 
=0 \ .
\label{eq_magnetic}
\ee
First of all, as one of the requirements of duality,
the magnetic theory possesses the same anomaly-free global
symmetries as the electric one
$
SU(N_f) \times SU(N_f) \times U(1)_B \times U(1)_R \ ,
$
which transform the dual quark superfields as 
$({\bar N_f},1,N_c/(N_f-N_c),N_c/N_f)$,  
the antiquark ones as 
$(1,N_f,-N_c/(N_f-N_c),N_c/N_f)$  and the gauge-singlet
superfields 
$M$ as $(N_f,{\bar N_f}, 0 , 2(N_f-N_c)/N_f)$.
The equations of the moduli space are   invariant under those symmetries; 
in
particular the invariance under $U(N_f)\times U(N_f)$ allows using the
general form of  solutions (\ref{vac_gene}), where $N_c$ is  replaced by
$N_f-N_c$. 
Then, the last equation in  (\ref{eq_magnetic}) implies that $a_i=0$,
\ie the moduli space of the magnetic theory  possesses  a baryonic 
solution  in the non-singular case, $\lambda \not = 0$~\footnote{As for the
electric case,  the baryonic branch consists of two baryonic orbits.  We
explicitly describe one of them.}.  Using the dilation invariance of the
equations one can get  the canonical solution on this orbit:
\be
q^i_{0,\alpha} = 
\left\{ 
\begin{array}{cl}
\delta^i_\alpha & \mbox{for } \  i=1\cdots N_c\ ;\\
0 & \mbox{for } \  i=N_f-N_c \cdots N_f\ ;
\end{array}
\right.
&&
{\tilde q}^{\tilde \imath}_{0,\tilde \alpha} = 0
\ .
\ee
We can now use this information to solve for $M$. It is easy to see that
\be
M &=&
\left(
\begin{array}{cc}
0& M_0\\
\end{array}
\right) \ ,
\ee
where $M_0$ is an $N_f\times N_c$ complex matrix.  We have identified the
regular part of the magnetic moduli space  as a set of  $SU(N_f) \times SU(N_f)
\times U(1)_B \times U(1)_R  \times SU(N_f-N_c)_G \times D$ orbits. Indeed the
magnetic vacua are obtained by the transitive action of  this symmetry group on
the canonical solutions.  The action of $U\in U(N_f)$ on the solutions is given
by  $q\to Uq$ and $M\to MU^{\dagger}$.  The stabilizer of the canonical value
of $q$ is  \be
H_m= SU(N_c)\times SU(N_f-N_c)_D \times SU(N_f) \times U(1)^2 \ ,
\ee
where $SU(N_c)$ is a subgroup of $SU(N_f)$; $SU(N_f-N_c)_D$ is the diagonal
combination of the gauge-group  and the flavour transformations; and the two
$U(1)$'s are combinations of the $U(1)_R$, the $U(1)_B$ and an element of the
Cartan subalgebra of $SU(N_f)$.  It is easy to see that the action of an element  
$h\in H_m$ of the stabilizer on the canonical field  $M=(0\ M_0)$ gives an element
$(0\ M'_0)$ in the canonical form.  The magnetic moduli space has a structure
of a fibred space.  After the extraction of  the gauge degrees of freedom of
the completely broken gauge group, the regular part has a base 
manifold  
\be
G_m/H_m=
{\frac{ SU(N_f) \times U(1)_R \times U(1)_B \times D}
{SU(N_c) \times SU(N_f-N_c)_D \times U(1)^2} }
\ ,
\label{Grassmanian}
\ee
whose real dimension is $2N_fN_c-2N_c^2+2$. 

The fibres are subspaces of ${\Bbb{C}}^{N_f\times N_c}$. More precisely  the
moduli space is parametrized by pairs $(U,M)$, where $U$ is a representative of
the $G_m/H_m$  and $M=(0\ M_0)U^{\dagger}$.  The corresponding vacuum
is given by $q=Uq_0$ and $M$.  
The action of the stabilizer $H_m$ leaves the base $G_m/H_m$ invariant  but
acts as an automorphism on the fibres.

Let us describe the base manifold.  As a homogeneous space it is characterized
by its tangent space  at the origin.  The tangent space is parametrized by
$N_f\times N_f$ matrices 
\be 
t(u) =
\left(
\begin{array} {cc}
0&u\\
-u^{\dagger}&0\\
\end{array}
\right) \ ,
\label{complex_structure}
\ee
where $u$ is a $(N_f-N_c)\times N_c$ complex matrix 
such that $U=e^{t(u)}$ are the elements of the coset $G_m/H_m$.  
This base manifold is a
complex manifold.  As the fibres are complex, this proves
that  the regular orbit of the magnetic moduli space is  a complex manifold. 

  Finally, the real dimension of this fibred
space is $4N_cN_f-2N^2_c+2$, in accordance with the expected dimension of the
electric moduli space and also with the number of massless states deduced
from the scalar mass matrix. Indeed, the magnetic theory
has $6 N_f^2 - 4 N_f N_c$ scalars.
Since the gauge group $SU(N_f-N_c)$ is completely broken, one has to exclude
$2(N_f-N_c)^2-2$ states in the vector multiplets.
Finally, the rank of the Hessian $F_{ij}$ matrix is
$2N_F (N_f-N_c)$, independently of the $M_0$ matrix elements.
This leave precisely $4N_f N_c -2 N_c^2 +2$ massless states. 

The singularities of the magnetic moduli space 
are obtained when the values of $q$ and $\tilde q$ are
$ q=\tilde q=0$. 
Hence any matrix $M$ is a solution of  
(\ref{eq_magnetic}).
The set of matrices $\{M\}$ is stratified by their rank $r$. 
In this letter, we concentrate on the baryonic orbits, and we shall
discuss the singularities corresponding to mesonic orbits in a forthcoming paper
\cite{Us}.

\vspace{.2cm}
It is appealing to try to understand the relation between 
the electric and magnetic
moduli spaces. We have seen that the  magnetic moduli space is parametrized by
the coordinates $u$ of the base manifold and the
coordinates along  the fibres $M_0$.  The following $N_c \times
(2N_f-N_c)$ complex matrix can be constructed: 
\be 
z_m=(u^T\ M_0^T)\ .
\label{z_m}
\ee
Consider the homographic action of the  group $U(N_f,N_f)$ on $z_m$, 
analogous to  (\ref{homographic}).
The magnetic moduli space is invariant
under this non-linear action  of $U(N_f,N_f)$. Indeed all the equations in 
(\ref{eq_magnetic}) are
invariant since $U(N_f,N_f)$ is an automorphism of the parameter space 
$z_m$ of 
their solutions. 
Consider the image of the origin $z_m=(0,0)$. As the action on $z_m$ is the 
same as (\ref{homographic}) on the electric baryonic orbit  
we know that the image of the
origin is the whole  baryonic orbit.   
Therefore the  baryonic branches  of the electric
and magnetic moduli spaces are   isomorphic as non-compact complex manifolds
if and only if the matrices (\ref{z_m}) are restricted to the $U(N_f,N_f)$ 
orbits.
The dualising map reads simply $ z \leftrightarrow z_m $. 
This identification is consistent with the fact that the subspace of
the $u$ submatrices is  given in both the electric and magnetic theories by the
coset (\ref{Grassmanian}).
The corresponding restriction of the matrices $M_0$ must result from 
the K\"ahler manifold for the meson fields $M$ associated to the
K\"ahler potential (\ref{K_magnetic}). 
The K\"ahler manifold satisfying these requirements is the homogeneous
space $U(N_f,N_f) / U(N_f) \times U(N_f)$.
The mesonic branches of the moduli space
close the baryonic one \cite{Us}.

We therefore conclude that the differential geometric description of 
the electric and magnetic moduli spaces shows an isomorphism provided that
the mesonic fields $M$ manifold is suitably defined. 
This isomorphism can be extended to the induced K\"ahlerian structure that we now
turn to discuss.


\vspace{.2cm}
Let us adopt the prescription given in \cite{ADS}. The induced K\"ahler potential is
deduced from the induced metric that one  can calculate by restricting the
K\"ahler potential  to the moduli space. Replacing the parametrization  of the
baryonic moduli space 
\be
\left(
\ba{c}
Q\\
{\tilde Q}^*
\ea
\right)_{\vert D^A=0}
\, = \, 
e^{t(z)}
\left(
\ba{c}
{\bold 1}_{N_c}\\
0
\ea
\right)\ ,
\ee
into (\ref{K_electric}), one obtains the induced K\"ahler potential
\be
{\cal K}_{eff}=
\tr ( Q^\dagger Q + {\tilde Q}^\dagger {\tilde Q})_{\vert D^A=0}
= \tr (e^{t(z)^{\dagger}}e^{t(z)}P_{N_c}) \ ,
\label{projector}
\ee
where $P_{N_c}$ projects onto the first $N_c$ components. 
This expression gives an explicit description of the low-energy theory. 

Let us now deal with the dual theory, where the parametrization of
the baryonic solutions of (\ref{eq_magnetic}) is
\be
q & = & e^{t(u)}q_0 \ ;\nonumber\\
{\tilde q} & = & 0 \ ;\nonumber\\
M & = & M_0 e^{t(u)  \dagger} \ .
\ee
The low-energy physics is entirely defined by the restriction of the 
K\"ahler potential (\ref{K_magnetic}) to these flat directions.
This yields
\be
{\cal K}_{eff}= \tr (e^{t(u)^{\dagger}}e^{t(u)}P_{N_f-N_c}) 
+ K ( e^{t(u)} M_0^{\dagger}   M_0 e^{t(u)^{\dagger}} ) \ ,
\label{magnetic_K}
\ee
where $P_{N_f-N_c}$ projects onto the first $N_f-N_c$ components. It is easily
checked from $z_m=(u^T\ M_0^T)$ that (\ref{projector}) and
(\ref{magnetic_K}), near the origin, exactly coincide with a Euclidean metrics. 
It should be noticed that in spite of the Minkowskian signature of the moduli
manifold, the metrics defined by (\ref{projector}) and (\ref{magnetic_K})
have Euclidean signature.

We have only obtained these results by looking at the scalar
sector of the theories. This approach to the low-energy 
effective theory is puzzling because the parametrization
of the solutions in terms of the moduli is not holomorphic. 
The flatness condition $D^A=0$ are not holomorphic (\cf the presence
of the K\"ahler function in (\ref{invariant}) ).
The $U(N_f,N_f)$ action is not analytic either.
Therfore we cannot directly extend our formalism onto a superfield one,
which would manifestly preserve supersymmetry.
This unsatisfactory aspect of our approach is now under investigation.

\vspace{.2cm}
The extension of this analysis to $SO(N_C)$ 
gauge supersymmetric theories \cite{dualitySO}
 with $N_f$  flavours of quarks transforming as
fundamental representations is straightforward \cite{Us}. Basically, the
non-compact symmetries $U(N_f,N_f)$ are replaced by their real subgroup
$Sp(2N_f, {\Bbb{R}})$ acting on $( \Ree Q^i_{\alpha} \ \Imm Q^i_{\alpha} )$.

\vspace{.2cm}
In conclusion,
we have shown the equivalence between the low-energy description of 
the electric and  magnetic $SU(N_c)$ supersymmetric gauge theories from a
more geometrical point of view. 
It would be of great interest to analyse the actions in terms of 
conformal field theories.

\section*{Acknowledgements}
We thank J.P. Derendinger, E. Dudas and R. Stora for helpful
and stimulating discussions.


\end{document}